\documentclass{article}
\usepackage{spconf,amsmath,amssymb,graphicx,hyperref}

\title{NVMOS: NON-VERBAL VOCALIZATION QUALITY ASSESSMENT IN SPEECH}
\name{Jialong Mai$^{1}$ \qquad Jinxin Ji$^{2}$ \qquad Xiaofen Xing$^{1,\ast}$ \qquad Wencui Liu$^{2}$ \qquad Xiangmin Xu$^{3}$\thanks{The NVMOS model has been released at \url{https://github.com/yongaifadian1/NVMOS}.}}
\address{$^{1}$South China University of Technology \\
$^{2}$Tongji University \\
$^{3}$Foshan University \\
$^{\ast}$Corresponding author}
\begin{document}
%
\maketitle
\begin{abstract}
Non-verbal vocalizations (NVs), such as laughter, sighs, and coughs, are important acoustic cues for emotion and intent. Existing speech quality assessment methods typically focus on overall naturalness, while non-verbal TTS evaluations mainly examine whether a target NV appears with the correct type and position. However, the perceptual quality of NV events themselves remains underexplored. To address this gap, we construct an NV-MOS dataset containing outputs from multiple NV-TTS systems and naturally occurring NV samples, with ratings collected from three acoustic experts on a perceptual quality scale. We further analyze audio-capable multimodal large language models such as Gemini and find clear inconsistencies between their scores and expert ratings. These results suggest that general-purpose multimodal models cannot reliably replace human judgments for NV quality assessment. We then propose NVMOS, to our knowledge the first model that can reliably predict the perceptual quality of NV events in speech. Experimental results show that, with a local NV-event focusing module, NVMOS reaches expert-level or stronger agreement with human MOS.
\end{abstract}

\section{Introduction}
\label{sec:intro}

Non-verbal vocalizations (NVs), such as laughter, sighs, and coughs, are
important acoustic cues for expressing emotion and intent in speech. As
generative speech systems move beyond intelligible and natural speech toward
more expressive and controllable communication~\cite{nsvtts2023,nonverbaltts2025,magictts2026}, the ability to produce appropriate NVs, and to
make them sound perceptually natural, becomes an important aspect of speech
generation quality.

Existing evaluations of NVs mainly focus on controllability~\cite{nvbench2026,nvbenchhyphen2026}. In non-verbal
text-to-speech (NV-TTS), recent benchmarks typically ask whether a target NV is
present, whether its category is correct, and whether it occurs at the expected
position~\cite{mnv17_2025}. These metrics indicate whether a system follows the control signal. Some NV
evaluations also include perceptual quality judgments, but they usually remain
at the stage of subjective MOS listening tests, making it difficult to scale the
evaluation to large numbers of generated samples. They do not answer a more
operational perceptual question: even if the target NV is generated, does it
sound natural, expressive, and smoothly integrated with the surrounding speech?

In parallel, general speech quality assessment and MOS prediction methods
usually estimate the overall naturalness or quality of an utterance~\cite{mosnet2019,nisqa2021}. They do not
explicitly model the perceptual quality of local NV events, making it difficult
to distinguish cases where the whole utterance is acceptable but the NV is
abrupt, or where the NV category is correct but the sound itself is unnatural.
This motivates a dedicated quality assessment target for non-verbal
vocalizations in speech.

To this end, we construct an NV-MOS dataset that contains synthetic outputs from
multiple NV-TTS systems and naturally occurring NV samples from real speech. The
dataset is designed to support both NV quality analysis for generated speech and
quality assessment of NV events in natural or wild speech. Each sample is rated
by three acoustic experts on a perceptual quality scale.

We further analyze whether audio-capable multimodal large language models, such
as Gemini~\cite{gemini2023}, can serve as automatic NV quality judges. Preliminary manual review
and subsequent numerical analysis show that such general-purpose models can
underestimate high-quality NVs, confuse whether an NV is present, misidentify NV
types, and produce explanations that are inconsistent with human perception.
These observations suggest that expert-aligned NV quality assessment requires a
specialized model.

Based on these findings, we propose NVMOS, to our knowledge the first automatic
model that can reliably predict the perceptual quality of NV events in speech.
NVMOS uses a local NV-event focusing module to emphasize time regions related to
the target NV and predicts quality scores aligned with expert judgments.

The main contributions of this work are:
\begin{itemize}
    \item We introduce NV-MOS, an expert-rated dataset for perceptual quality
    assessment of non-verbal vocalizations in speech, covering both synthetic
    NV-TTS outputs and naturally occurring NV samples.
    \item We analyze audio-capable multimodal large language models such as
    Gemini and show that they are not reliable substitutes for expert NV
    quality judgments.
    \item We propose NVMOS, the first reliable NV quality prediction model for
    speech, and show that its local event-focused design achieves expert-level
    or stronger agreement with human MOS.
\end{itemize}

\section{NV-MOS Dataset}
\label{sec:dataset}

\subsection{Data Sources}
\label{ssec:data_sources}

NV-MOS contains both synthetic speech samples from multiple NV-TTS systems
~\cite{smiipnv2025,chatterboxtts2025,dia2025} and natural speech samples from
open-source datasets~\cite{mnv17_2025,nvspeech2025,nonverbalspeech38k2025} and
a natural speech corpus~\cite{emilia2024}. The target texts for synthetic
samples are generated by a large language model~\cite{gemini2023}, with target
NV marks distributed across the beginning, middle, and end positions of
utterances; we also train an extra NV-TTS system to broaden the synthetic NV
coverage. After balancing different NV categories, the final dataset contains
7,784 samples with about 9.51 hours of speech, including 2,655 synthetic samples
and 5,129 natural samples. It covers 16 NV categories and, to the best of our
knowledge, spans the NV support range of existing NV-TTS systems.

\subsection{Annotation Protocol}
\label{ssec:annotation}

The annotation target of NV-MOS is the perceptual quality of a non-verbal
vocalization event. Raters listen to each audio sample with the corresponding
text containing an NV mark, and assign a discrete quality score to the marked
NV event. The rating mainly considers whether the target NV appears at the
marked position, whether the NV itself sounds natural and clear with a
reasonable acoustic form, and whether it is smoothly connected with the
preceding and following speech.

We use a 0--5 rating scale, where 5 denotes a highly natural and expressive
NVC, 4 a generally natural NVC, 3 average naturalness with some synthetic
feeling, 2 a stiff or poorly connected NVC, 1 a clearly unnatural NVC, and 0 a
barely audible or absent NVC.


Each sample is rated by three PhD-level acoustic experts. For samples with
inconsistent ratings, we first perform a review and resolution step, and then
use the average of the three expert scores as the final continuous label. In
the following experiments, this averaged expert score is used as the main
supervision signal, and model predictions are evaluated by both error metrics
and correlation with expert judgments.

\subsection{Data Split}
\label{ssec:data_split}

The dataset is split into training, validation, and test sets with an
approximately 9:0.5:0.5 ratio. The training set contains 7,006 samples, while
the validation and test sets contain 389 samples each.

During splitting, for natural speech, samples from the same speaker appear in
only one split, so that the model does not see the same speaker in training and
testing. For synthetic speech, samples from the same synthesis system are also
kept in the same split as much as possible, reducing synthetic-pattern leakage
between training and test sets. In addition, we control the proportions of
different data sources and NV categories across the training, validation, and
test sets, so that the three splits have similar compositions.

\section{Are Multimodal Judges Reliable for NV Quality?}
\label{sec:judge}

As audio understanding is increasingly integrated into general-purpose
multimodal large language models, a natural question is whether such models can
directly replace human experts for NV quality assessment. Compared with
traditional MOS prediction models, LLM judges are attractive because they do
not require task-specific training, can take both audio and textual context as
input, and can provide natural-language explanations. Before building a
dedicated NVMOS model, we therefore examine the reliability of general audio
multimodal models as NV quality judges.

NVBench~\cite{nvbench2026} first introduced an LLM-based judge protocol
for NV generation evaluation, where an audio multimodal model is used to judge
whether a target NV is present and the quality of the generated NV. Our LLM
judge protocol follows the same general idea, focusing on the perceptual
quality of the generated target NV event. Given an audio sample and its text
with an NV mark, the judge is asked to assess whether the target NV appears,
whether the sound itself is natural and clear, and whether it is smoothly
connected with the surrounding speech. The judge then outputs a 0--5 quality score, where
0 denotes a barely audible or absent NV and 5 denotes a highly natural and
expressive NV. This setting is aligned with the NV-MOS expert annotation
protocol, allowing LLM judge scores to be directly compared with expert MOS.

We evaluate five LLM judge settings, including Gemini 2.5 Pro, Gemini 3
Flash, MOSS-Audio 4B, MOSS-Audio 8B, and Qwen-Omni 30B. For each model, we
compute the agreement between its predicted scores and expert MOS on the
NV-MOS test set. As shown in
Table~\ref{tab:llm_judge_corr}, Gemini 3 Flash obtains the highest correlation
with expert MOS.
We primarily report Pearson, Spearman, and Kendall correlations to measure
linear agreement, rank agreement, and pairwise ordering consistency,
respectively. MAE is included as a secondary calibration indicator.

\begin{table}[t]
\centering
\caption{Agreement between LLM judges and expert MOS on the NV-MOS test set.}
\label{tab:llm_judge_corr}
\scriptsize
\begin{tabular}{lrrrr}
\hline
Judge & Pearson & Spearman & Kendall & MAE \\
\hline
Gemini 3 Flash & 0.468 & 0.453 & 0.377 & 1.036 \\
Gemini 2.5 Pro$^\dagger$ & 0.381 & 0.372 & 0.302 & 1.221 \\
MOSS-Audio 8B & 0.173 & 0.113 & 0.095 & 1.393 \\
MOSS-Audio 4B & 0.002 & -0.030 & -0.027 & 1.938 \\
Qwen-Omni 30B & -0.049 & -0.094 & -0.082 & 1.789 \\
\hline
\end{tabular}
\vspace{1mm}

\footnotesize{$^\dagger$Gemini 2.5 Pro is the LLM judge used in NVBench.}
\end{table}

Even the best LLM judge, Gemini 3 Flash, is still far below the inter-expert
agreement level: its Pearson correlation with expert MOS is 0.468, while the
pairwise expert Pearson correlations range from 0.589 to 0.699.

Manual review further shows that LLM judge errors involve fundamental auditory
judgment errors.
First, LLM judges can underestimate high-quality NVs. For example, natural,
clear, and contextually appropriate sighs, lip smacks, or sniffles may receive
low scores, with the explanation claiming that the target NV is absent. Second,
there are also cases where the target NV is absent but the model still assigns
a high score. Third, LLM judges may confuse NV categories, such as interpreting
a sneeze as laughter or mistaking another vocal event for the target category.

These observations suggest that NV quality assessment is different from general
audio understanding or utterance-level speech quality assessment. The target NV
is often a local, short, low-energy, or ambiguous acoustic event. Its quality
depends not only on whether the event is present, but also on its intensity,
duration, expressive function, and connection with the surrounding speech.
Although general multimodal models have some audio
understanding ability, they are not calibrated for these local NV perceptual
dimensions. As a result, they may rely on global speech naturalness, textual
semantics, or category priors, rather than truly focusing on the target NV
event itself.

\section{NVMOS: Local Event-Focused NV Quality Assessment}
\label{sec:model}

\begin{figure*}[t]
    \centering
    \includegraphics[width=0.9\textwidth]{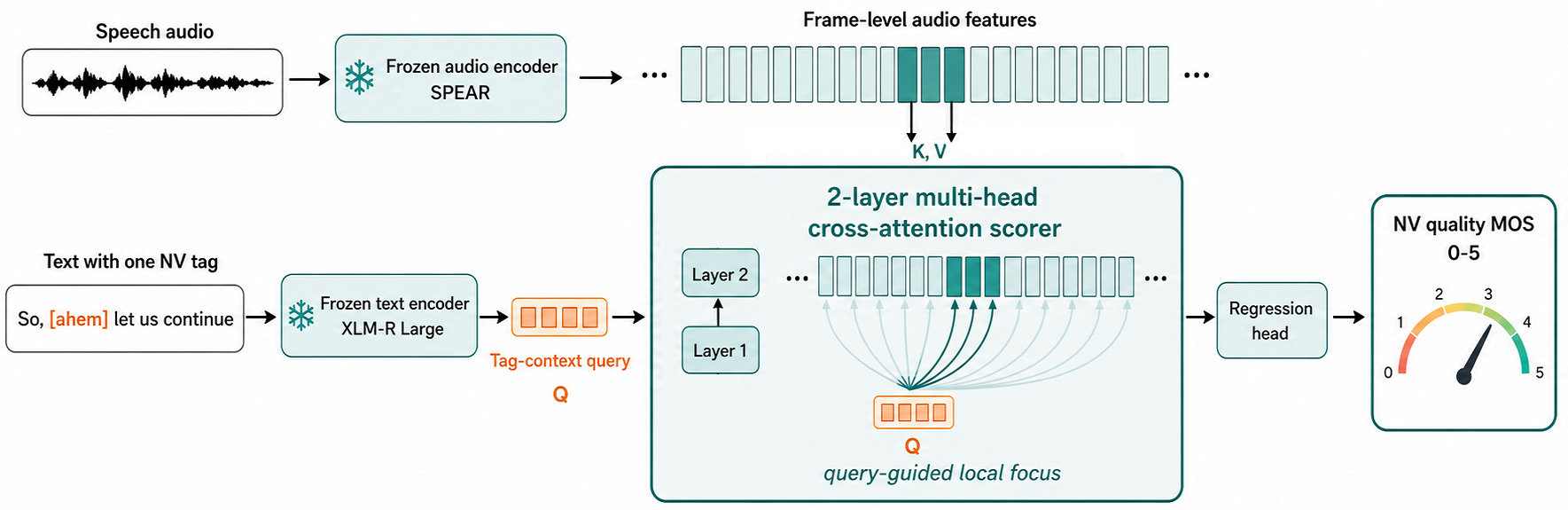}
    \caption{Overview of the proposed NVMOS framework. The marked text provides a tag-centered query, which guides cross-attention over frame-level speech representations for local NV quality prediction.}
    \label{fig:nvmos_arch}
\end{figure*}

Unlike conventional MOS prediction, the input to NVMOS contains both the audio
signal and the corresponding text with an explicit NV mark. The goal is not to
score the utterance as a whole, but to predict the perceptual quality of the
target NV event while still considering whether the event is coherent with its
surrounding speech. The overall architecture is shown in Figure~\ref{fig:nvmos_arch}. We therefore formulate NV-MOS prediction as a text-queried
audio quality assessment problem.

\subsection{Text-Queried Local Focusing}
\label{ssec:local_focusing}

Given an utterance audio $a$ and its marked text $t$, we first extract a
frame-level speech representation
$\mathbf{A}=[\mathbf{a}_1,\ldots,\mathbf{a}_T]\in\mathbb{R}^{T\times d_a}$
from a speech encoder. In this work, we mainly study WavLM Large
representations~\cite{wavlm2022} and SPEAR representations~\cite{spear2025}.
We keep the temporal feature sequence instead of using an utterance-level pooled
vector, so that the downstream model can attend to local regions that are
relevant to the target NV.

For the text side, we encode the marked text with XLM-R
Large~\cite{conneau2020xlmr}, which is suitable for the bilingual English and
Chinese text in NV-MOS. The target NV tag is already enclosed in brackets in
the text, such as \texttt{[ahem]}. We locate the character span of the tag
inside the brackets and use the tokenizer offset mapping to find the
corresponding token span. The hidden states of these tag tokens are averaged to
form a compact query representation:
\begin{equation}
    \mathbf{q} = \frac{1}{|\mathcal{I}_{\mathrm{tag}}|}
    \sum_{i\in\mathcal{I}_{\mathrm{tag}}}\mathbf{h}_i,
\end{equation}
where $\mathcal{I}_{\mathrm{tag}}$ denotes the token indices overlapping with
the NV tag span. The resulting vector is then projected to the model dimension.
This design makes the text query explicitly centered on the marked NV, instead
of relying on a sentence-level text embedding where the target tag may be
diluted.

The core component of NVMOS is a text-queried local focusing module. The text
query is used as the query sequence, and the frame-level audio features are
used as keys and values in a stack of cross-attention layers:
\begin{equation}
    \mathbf{z}^{(l)} =
    \mathrm{CrossAttn}(\mathbf{z}^{(l-1)}, \mathbf{A}, \mathbf{A}),
\end{equation}
where $\mathbf{z}^{(0)}$ is initialized from the projected tag-context query.
This module is implemented with standard multi-head attention
~\cite{vaswani2017attention}. Although no hard NV boundary is given, the query
provides a soft target for the downstream model: the model can learn to assign
higher attention to audio frames that are useful for judging the marked NV and
its local connection to neighboring speech.

After two cross-attention layers, we average the output query states and pass
the result to a feed-forward regression head:
\begin{equation}
    \hat{y} = f_{\mathrm{reg}}(\mathrm{MeanPool}(\mathbf{z}^{(L)})).
\end{equation}
The model is trained with a robust regression loss between the predicted score
$\hat{y}$ and the expert MOS target $y$.

\section{Experiments}
\label{sec:experiments}

\subsection{Experimental Setup}
\label{ssec:exp_setup}

We evaluate NVMOS on the NV-MOS train, validation, and test splits described in
Section~\ref{ssec:data_split}, using Pearson correlation as the primary metric
and reporting Spearman correlation, Kendall's tau-b, and MAE for comparison.
The downstream scorer uses two cross-attention layers, eight attention heads,
hidden size 256, FFN size 1024, and dropout 0.1, and is trained for 10 epochs
with AdamW, learning rate $10^{-4}$, weight decay $10^{-2}$, batch size 8,
Smooth L1 loss, and gradient clipping at 1.0.

%

\subsection{Main Results}
\label{ssec:main_results}

Table~\ref{tab:nvmos_main_results} reports NVMOS results using WavLM Large
layer 7 and SPEAR Large layer 9 as speech representations. With WavLM, NVMOS
obtains a test Pearson correlation of 0.697, which is comparable to the
inter-expert agreement range. With SPEAR, NVMOS obtains a similar test Pearson
correlation of 0.690 and also clearly
outperforms the zero-shot LLM judges in
Table~\ref{tab:llm_judge_corr}. These results indicate that the proposed
text-queried local focusing model captures expert NV quality preferences more
reliably than general-purpose multimodal judges.
Considering the three correlation metrics together, NVMOS achieves stronger
agreement with expert MOS on the test set than the pairwise agreement among
experts, showing that it can reliably assess NV quality in speech.

\begin{table}[t]
\centering
\caption{Representative NVMOS results with text-queried cross-attention.}
\label{tab:nvmos_main_results}
\scriptsize
\setlength{\tabcolsep}{2.5pt}
\begin{tabular}{lrrrrr}
\hline
Feature & L & Pearson & Spearman & Kendall & MAE \\
\hline
WavLM Large & 7 & 0.697 & 0.657 & 0.518 & 0.837 \\
SPEAR Large & 9 & 0.690 & 0.664 & 0.524 & 0.791 \\
\hline
\end{tabular}
\end{table}

\subsection{Ablation Study}
\label{ssec:ablation}

We conduct an ablation study on SPEAR Large layer 9 to examine whether the
text input is useful and, more specifically, whether the model benefits from a
tag-centered text query. All systems use the same frame-level audio features,
training split, and cross-attention scorer. The audio-only system replaces the
text query with a learned query vector. The full-text mean system uses an
attention-mask weighted mean of all XLM-R token states as a single query. The
full-text token system keeps the whole XLM-R token sequence as multiple
queries, cross-attends them to the audio frames, and then mean-pools the
updated token states for prediction. The proposed system uses only the
tag-context query centered on the marked NV.

\begin{table}[t]
\centering
\caption{Text-query ablation on SPEAR Large layer 9.}
\label{tab:ablation}
\scriptsize
\setlength{\tabcolsep}{2.5pt}
\begin{tabular}{lrrrr}
\hline
System & Pearson & Spearman & Kendall & MAE \\
\hline
Audio only & 0.581 & 0.557 & 0.434 & 0.896 \\
Full-text mean & 0.644 & 0.619 & 0.485 & 0.828 \\
Full-text tokens & 0.635 & 0.611 & 0.480 & 0.871 \\
Tag-context query & 0.690 & 0.664 & 0.524 & 0.791 \\
\hline
\end{tabular}
\end{table}

The results show that using the full sentence as a generic text query brings
little benefit over the audio-only baseline, regardless of whether the sentence
is compressed by mean pooling or kept as token-level queries. In contrast, the
tag-context query substantially improves the correlation metrics as well as
score calibration. This
supports the main design choice of NVMOS: the text input should not merely
provide global sentence semantics, but should explicitly guide the scorer toward
the marked NV event.

\section{Conclusion}
\label{sec:conclusion}

We presented NV-MOS, an expert-rated dataset and modeling framework for
perceptual quality assessment of non-verbal vocalizations in speech. Our
analysis shows that general audio-capable LLM judges remain poorly aligned with
expert NV quality ratings, motivating a dedicated task-specific predictor.
NVMOS addresses this problem with a text-queried cross-attention scorer that
uses the marked NV tag to focus on relevant frame-level audio regions.
Experimental results show that, with its local NV-event focusing module, NVMOS
achieves expert-level or stronger agreement with human MOS, making it a reliable
model for assessing NV quality in speech.

\bibliographystyle{IEEEbib}
\bibliography{strings,refs}

@inproceedings{nsvtts2023,
  author    = {Zhang, Haitong and Yu, Xinyuan and Lin, Yue},
  title     = {{NSV-TTS}: Non-Speech Vocalization Modeling and Transfer in Emotional Text-to-Speech},
  booktitle = {Proc. IEEE International Conference on Acoustics, Speech and Signal Processing (ICASSP)},
  year      = {2023},
  pages     = {1--5},
  doi       = {10.1109/ICASSP49357.2023.10096033}
}

@article{nonverbaltts2025,
  author        = {Borisov, Maksim and Spirin, Egor and Diatlova, Daria},
  title         = {{NonverbalTTS}: A Public English Corpus of Text-Aligned Nonverbal Vocalizations with Emotion Annotations for Text-to-Speech},
  journal       = {arXiv preprint arXiv:2507.13155},
  year          = {2025},
  eprint        = {2507.13155},
  archivePrefix = {arXiv},
  primaryClass  = {cs.LG},
  doi           = {10.48550/arXiv.2507.13155}
}

@article{magictts2026,
  author        = {Mai, Jialong and Xing, Xiaofen and Xu, Xiangmin},
  title         = {{MAGIC-TTS}: Fine-Grained Controllable Speech Synthesis with Explicit Local Duration and Pause Control},
  journal       = {arXiv preprint arXiv:2604.21164},
  year          = {2026},
  eprint        = {2604.21164},
  archivePrefix = {arXiv},
  primaryClass  = {cs.SD}
}

@article{nvbench2026,
  author        = {Xue, Liumeng and others},
  title         = {{NVBench}: A Benchmark for Speech Synthesis with Non-Verbal Vocalizations},
  journal       = {arXiv preprint arXiv:2604.16211},
  year          = {2026},
  eprint        = {2604.16211},
  archivePrefix = {arXiv},
  primaryClass  = {cs.SD}
}

@article{nvbenchhyphen2026,
  author        = {Ni, Qinke and Liao, Huan and Chen, Dekun and Wang, Yuxiang and Wu, Zhizheng},
  title         = {{NV-Bench}: Benchmark of Nonverbal Vocalization Synthesis for Expressive Text-to-Speech Generation},
  journal       = {arXiv preprint arXiv:2603.15352},
  year          = {2026},
  eprint        = {2603.15352},
  archivePrefix = {arXiv},
  primaryClass  = {cs.SD}
}

@article{mnv17_2025,
  author        = {Mai, Jialong and Ji, Jinxin and Xing, Xiaofen and Yang, Chen and Chen, Weidong and Xing, Jingyuan and Xu, Xiangmin},
  title         = {{MNV-17}: A High-Quality Performative Mandarin Dataset for Nonverbal Vocalization Recognition in Speech},
  journal       = {arXiv preprint arXiv:2509.18196},
  year          = {2025},
  eprint        = {2509.18196},
  archivePrefix = {arXiv},
  primaryClass  = {cs.SD}
}

@inproceedings{mosnet2019,
  author    = {Lo, Chen-Chou and Fu, Szu-Wei and Huang, Wen-Chin and Wang, Xin and Yamagishi, Junichi and Tsao, Yu and Wang, Hsin-Min},
  title     = {{MOSNet}: Deep Learning-Based Objective Assessment for Voice Conversion},
  booktitle = {Proc. Interspeech},
  year      = {2019},
  pages     = {1541--1545},
  doi       = {10.21437/Interspeech.2019-2003}
}

@inproceedings{nisqa2021,
  author    = {Mittag, Gabriel and Naderi, Babak and Chehadi, Assmaa and Moeller, Sebastian},
  title     = {{NISQA}: A Deep CNN-Self-Attention Model for Multidimensional Speech Quality Prediction with Crowdsourced Datasets},
  booktitle = {Proc. Interspeech},
  year      = {2021},
  pages     = {2127--2131},
  doi       = {10.21437/Interspeech.2021-299}
}

@article{wavlm2022,
  author        = {Chen, Sanyuan and others},
  title         = {{WavLM}: Large-Scale Self-Supervised Pre-Training for Full Stack Speech Processing},
  journal       = {IEEE Journal of Selected Topics in Signal Processing},
  year          = {2022},
  volume        = {16},
  number        = {6},
  pages         = {1505--1518},
  doi           = {10.1109/JSTSP.2022.3188113}
}

@article{spear2025,
  author        = {Yang, Xiaoyu and Yang, Yifan and Jin, Zengrui and Cui, Ziyun and Wu, Wen and Li, Baoxiang and Zhang, Chao and Woodland, Phil},
  title         = {{SPEAR}: A Unified {SSL} Framework for Learning Speech and Audio Representations},
  journal       = {arXiv preprint arXiv:2510.25955},
  year          = {2025},
  eprint        = {2510.25955},
  archivePrefix = {arXiv},
  primaryClass  = {cs.SD}
}

@inproceedings{conneau2020xlmr,
  author    = {Conneau, Alexis and others},
  title     = {Unsupervised Cross-lingual Representation Learning at Scale},
  booktitle = {Proc. ACL},
  year      = {2020},
  pages     = {8440--8451},
  doi       = {10.18653/v1/2020.acl-main.747}
}

@inproceedings{vaswani2017attention,
  author    = {Vaswani, Ashish and Shazeer, Noam and Parmar, Niki and Uszkoreit, Jakob and Jones, Llion and Gomez, Aidan N. and Kaiser, Lukasz and Polosukhin, Illia},
  title     = {Attention Is All You Need},
  booktitle = {Advances in Neural Information Processing Systems},
  year      = {2017},
  pages     = {5998--6008}
}

@article{gemini2023,
  author        = {{Gemini Team}},
  title         = {Gemini: A Family of Highly Capable Multimodal Models},
  journal       = {arXiv preprint arXiv:2312.11805},
  year          = {2023},
  eprint        = {2312.11805},
  archivePrefix = {arXiv},
  primaryClass  = {cs.CL}
}

@inproceedings{smiipnv2025,
  author    = {Wu, Zhuojun and Liu, Dong and Liu, Juan and Wang, Yechen and Li, Linxi and Jin, Liwei and Bu, Hui and Zhang, Pengyuan and Li, Ming},
  title     = {{SMIIP-NV}: A Multi-Annotation Non-Verbal Expressive Speech Corpus in Mandarin for LLM-Based Speech Synthesis},
  booktitle = {Proceedings of the 33rd ACM International Conference on Multimedia (MM '25)},
  year      = {2025},
  address   = {Dublin, Ireland},
  pages     = {1--7},
  doi       = {10.1145/3746027.3758312}
}

@misc{chatterboxtts2025,
  author       = {{Resemble AI}},
  title        = {{Chatterbox-TTS}},
  year         = {2025},
  howpublished = {GitHub repository},
  note         = {Available at \url{https://github.com/resemble-ai/chatterbox}}
}

@misc{dia2025,
  author       = {{Nari Labs}},
  title        = {{Dia}: A Text-to-Speech Model for Dialogue Generation},
  year         = {2025},
  howpublished = {GitHub repository},
  note         = {Available at \url{https://github.com/nari-labs/dia}}
}

@article{nvspeech2025,
  author        = {Liao, Huan and Ni, Qinke and Wang, Yuancheng and Lu, Yiheng and Zhan, Haoyue and Xie, Pengyuan and Zhang, Qiang and Wu, Zhizheng},
  title         = {{NVSpeech}: An Integrated and Scalable Pipeline for Human-Like Speech Modeling with Paralinguistic Vocalizations},
  journal       = {arXiv preprint arXiv:2508.04195},
  year          = {2025},
  eprint        = {2508.04195},
  archivePrefix = {arXiv},
  primaryClass  = {cs.SD},
  doi           = {10.48550/arXiv.2508.04195}
}

@article{nonverbalspeech38k2025,
  author        = {Ye, Runchuan and Zhou, Yixuan and Yu, Renjie and Lin, Zijian and Li, Kehan and Li, Xiang and Liu, Xin and Zeng, Guoyang and Wu, Zhiyong},
  title         = {A Scalable Pipeline for Enabling Non-Verbal Speech Generation and Understanding},
  journal       = {arXiv preprint arXiv:2508.05385},
  year          = {2025},
  eprint        = {2508.05385},
  archivePrefix = {arXiv},
  primaryClass  = {cs.SD},
  doi           = {10.48550/arXiv.2508.05385}
}

@article{emilia2024,
  author        = {He, Haorui and others},
  title         = {{Emilia}: An Extensive, Multilingual, and Diverse Speech Dataset for Large-Scale Speech Generation},
  journal       = {arXiv preprint arXiv:2407.05361},
  year          = {2024},
  eprint        = {2407.05361},
  archivePrefix = {arXiv},
  primaryClass  = {eess.AS},
  doi           = {10.48550/arXiv.2407.05361}
}

\end{document}